\documentclass[twocolumn,prl,showpacs,preprintnumbers,amsmath,amssymb]{revtex4}
\usepackage{color}
\usepackage{bm, graphicx, amsmath}
\usepackage[mathscr]{eucal}
\newcommand{\two}[2]{\begin{array}{c}\\[-1.5em]\scriptstyle #1\\[-.5em] \scriptstyle #2\end{array}}
\newcommand{\bb}{b}
\newcommand{\w}{w}
\newcommand{\ket}[1]{|#1\rangle}
\newtheorem{theo}{Theorem}

\newtheorem{theorem}{Theorem}
\newtheorem{lemma}[theorem]{Lemma}

\begin{document}

\title{Duality games and operational duality relations}
\author{Emilio Bagan$^{1}$, John Calsamiglia$^{1}$, J\'anos A. Bergou$^{2,3}$, and Mark Hillery$^{2,3}$}
\affiliation{$^{1}$F\'{i}sica Te\`{o}rica: Informaci\'{o} i Fen\`{o}mens Qu\`antics, Universitat Aut\`{o}noma de Barcelona, 08193 Bellaterra (Barcelona), Spain \\
$^{2}$Department of Physics and Astronomy, Hunter College of the City University of New York, 695 Park Avenue, New York, NY 10065 USA \\ 
$^{3}$Graduate Center of the City University of New York, 365 Fifth Avenue, New York, NY 10016}

\begin{abstract}
We give operational meaning to wave-particle duality in terms of discrimination games. Duality arises as a constraint on the probability of winning these games.  The games are played with the aid of an $n$-port interferometer, and involve 3 parties, Alice and Bob, who cooperate, and the House, who supervises the game.  In one game called \emph{ways} they attempt to determine the path of a particle in the interferometer.  In another, called \emph{phases}, they attempt to determine which set of known phases have been applied to the different paths.  The House determines which game is to be played by flipping a coin.  We find a tight wave-particle duality relation that allows us to relate the probabilities of winning these games, and use it to find an upper bound on the probability of winning the combined game.  This procedure allows us to describe wave-particle duality in terms of discrimination probabilities.
\end{abstract}


\maketitle

Quantum coherence has become an active area of research since its first treatment as a resource theory \cite{baumgratz}.  In this theory, one first chooses a basis and then specifies the set of incoherent states as those that are diagonal with respect to the chosen basis.  One also defines incoherent operations, which can be done in a number of different ways~\cite{baumgratz,winter,yadin,chitambar,marvian}, but a minimum requirement is that an incoherent operation not create coherence, i.e., that it maps the set of incoherent states into itself.  It is possible to quantify the coherence of a state by its distance from the set of incoherent states for some appropriate distance measure.  In \cite{baumgratz}, measures of coherence using the $\ell_{1}$ norm and the relative entropy were defined. Robustness of coherence is  yet another such measure recently introduced in~\cite{pianiadessoprl,pianiadessopra}, and shown to be a lower bound to the $\ell_1$-norm of coherence~\cite{baumgratz,winter}. The two measures were also shown to be equivalent for certain classes of states. At variance with other measures of coherence, robustness of coherence is both operational and observable.  A review of the current status of the study of quantum coherence can be found in~\cite{streltsov}.

The work on the quantification of coherence has led to a revival of studies of wave-particle duality.  A particle going through an interferometer can take any number of paths and eventually produce an interference pattern.  There is a trade-off in how much information one has about the path, and the strength of the interference pattern.  In the case of two paths, it was possible to quantify this relation by using the visibility of the interference pattern as a measure of the wave-like properties of the system \cite{wootters,greenberger,jaeger,englert}.  These relations take the form
\begin{equation}
\label{duality1}
D^{2}+V^{2}\leq 1 ,
\end{equation}
where $D$ is a measure of the path information and $V$ is the visibility. %the visibility $V$ quantifies the wave-like aspects of the particle.  
In the earlier studies the path information was related to the probabilities, derived from the state of the particle inside the interferometer, that the particle would take a particular path. In a seminal paper, Englert introduced path detectors, whose state changes if a particle passes through them, and he used the distinguishability of the detector states as a measure of the path information \cite{englert}.  These studies were extended to more than two paths using different measures of the wave properties of the system by a number of authors~\mbox{\cite{jaeger,durr,bimonte,englert2,jakob,englert3}}.  A different type of wave-particle duality relation using information quantities, which is not of the form given in Eq.~(\ref{duality1}), was derived by Angelo and Ribeiro \cite{angelo}.  With the advent of coherence measures, which are natural candidates for measures of the wave-like properties of a quantum system, additional duality relations were derived.  The first was by Bera, \emph{et al}.\ and used the $\ell_{1}$ coherence measure for the wave properties and an upper bound for the probability of successfully unambiguously distinguishing the path detector states for the path information~\cite{pati}.  In~\cite{bagan} two duality relations were derived.  The first made use of the~$\ell_{1}$ coherence measure and the probability of successfully distinguishing the detector states by using minimum-error discrimination.  The second used the entropic coherence measure and the mutual information between the path-detector states and any measurement to distinguish them.  All of the duality relations derived from the coherence measures are for $n$ paths.

\begin{figure}[htbp]
\begin{center}
\includegraphics[width=12em]{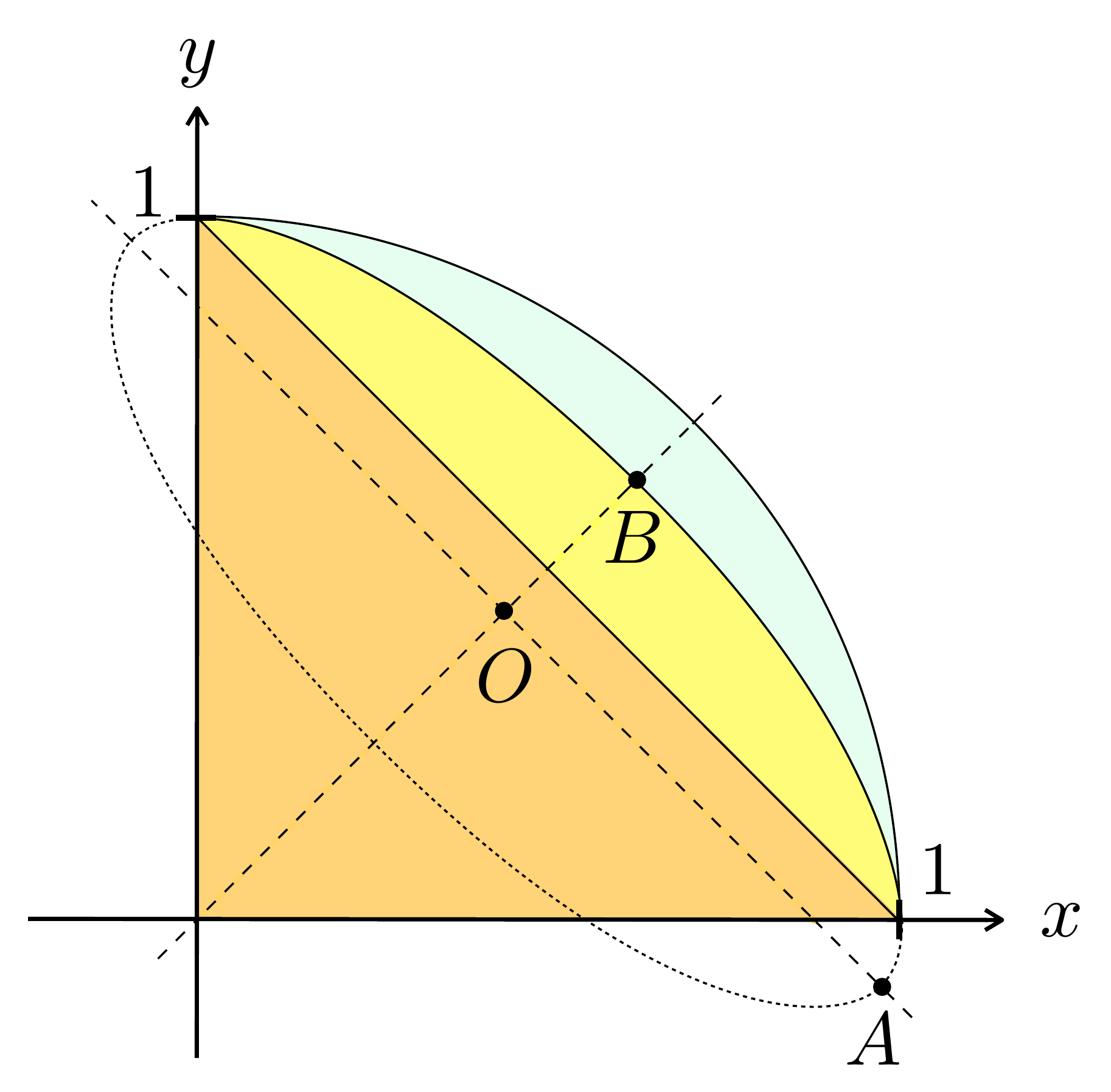}
\caption{Various regions involved in Theorem~\ref{t:2}, for $n=2$  (union of pale blue, yellow and orange regions), $n=9$~(union of yellow and orange regions), and $n\to\infty$  (orange region). The region defined by the bound in~\cite{bagan} coincides with the~\mbox{$n=2$} region.}
\label{f:2}
\end{center}
\end{figure}

In this letter, %we combine variations of the games discussed in the preceeding paragraphs into a single game to 
we
give an operational meaning to wave-particle duality entirely based on discrimination tasks. Discrimination is arguably the most fundamental information theory primitive. Thus, by doing so, we provide a very intuitive, robust and precise interpretation of duality.  We present a %tight inequality that supersedes/improves on the duality relation given in~\cite{bagan}.  %We will show that quantum mechanics sets a severe limit to the probability of winning this game. This limit is just a different facet of wave-particle duality, as will become clear. 
%The result follows from a tight bound on the $\ell_1$ coherence measure, which we call Theorem~\ref{t:1}. In the second part of the letter,  we use the theorem to introduce a novel wave-particle duality relation that generalizes that introduced in~\cite{bagan}. 
novel duality relation (Theorem~\ref{t:2}), much stronger than that given in~\cite{bagan}.
It defines a convex region on a plane given essentially by the success probability of phase discrimination  ---or alternatively, by the~$\ell_1$  coherence measure--- and the success probability~of path discrimination (Fig.~\ref{f:2}). Any physical  $n$-port interferometric experiment corresponds (or can be mapped) to a point in this region. For~$n=2$, the region is nothing but the positive quadrant of a circle, just like the region defined by Eq.~(\ref{duality1}). As $n$ becomes larger,  the shape of the region approaches a right triangle shape, which corresponds to a linear, rather than quadratic, duality relation. At variance with previous duality relations, ours is tight, and we identify input states and setups that define the boundary of the physical region. It is, therefore, the tightest duality relation involving the aforementioned quantities that one can write. 
%
%
%The duality relation stems from a novel inequality that is interesting on its own right, as it relatesÉ
%
The result follows from an attainable bound on the $\ell_1$ coherence measure, which we call Lemma~\ref{l:1}. Weaker, but simpler, duality relations can be derived from Lemma~\ref{l:1}. %our main inequality. 
In particular, before deriving Theorem~\ref{t:2}, %we give a linear such relation 
we give such a relation which is linear
(Theorem~\ref{t:1}). 

Our results are presented as a game, which consists of two parts, both of which are games in their own right.  There is a history of using games to analyze coherence and path information.  In~\cite{coles1} it was shown that wave particle duality relations of the form of Eq.~(\ref{duality1}) are equivalent to a formulation of the uncertainty principle in terms of entropies. Two games were used in the derivation: a ``particle" game and a ``wave" game. The setting for both is a two-path interferometer with ancillary systems attached.  In the first game, Alice (the player) can try to determine which path a photon took, and in the second, she tries to determine whether a phase of $\phi_{0}$ or $\phi_{0}+\pi$ was applied to the photon in one arm of the interferometer.  Alice's ability to win both games is limited by an entropic uncertainty relation that was proved in \cite{coles1}, from which the duality relations follow. %the authors show that bounds of the form of Eq.~(\ref{duality1}) can be derived.  
This was generalized to an $n$-path interferometer in \cite{coles2}. For path information the authors used the success probability of minimum error discrimination, as in~\cite{bagan}, and visibility to quantify the wave-like behavior of the particle in the interferometer.   In \cite{pianiadessoprl} the probability of winning a game to determine phase information was related to the robustness of coherence.%a coherence measure called the robustness of coherence.

Our formulation in terms of a game and discrimination probabilities enables us to formalize wave-particle duality in an abstract, mathematically precise, and model independent way as restrictions on the probabilities of winning these games.  
We begin with a discussion of the games. %For pedagogical reasons we first discuss the games. %and with the aid of two games and a combination of these games ---along the line of~\cite{coles1,coles2}--- that we believe have interest of their own. 

%Robustness of coherence (ROC) is an operational and observable measure of coherence recently introduced in~\cite{pianiadessoprl} and shown to be a lower bound to the $\ell_1$-norm of coherence~\cite{baumgratz,winter}. The two measures are also shown to be equivalent for certain classes of states. The operational interpretation of the ROC was elegantly presented with the aid of a game where the payer's (Alice's) task is to identify what  phases out of a given set were  imprinted to a state~$\rho$ in its passage through a black box. The phases are applied incoherently in some particular basis, which we will refer to as the path basis for obvious reasons, so states that are incoherent in this basis are useless for the game. Each set of phases, along with their corresponding prior probabilities of application, defines a possible game.  The authors showed that the player's maximum probability of success over all such games yields the ROC of $\rho$ through a very simple expression. So, ROC measures how useful~a state is when it comes to discriminate phases applied incoherently. 

%
\begin{figure}[htbp]
\begin{center}
\includegraphics[width=24em]{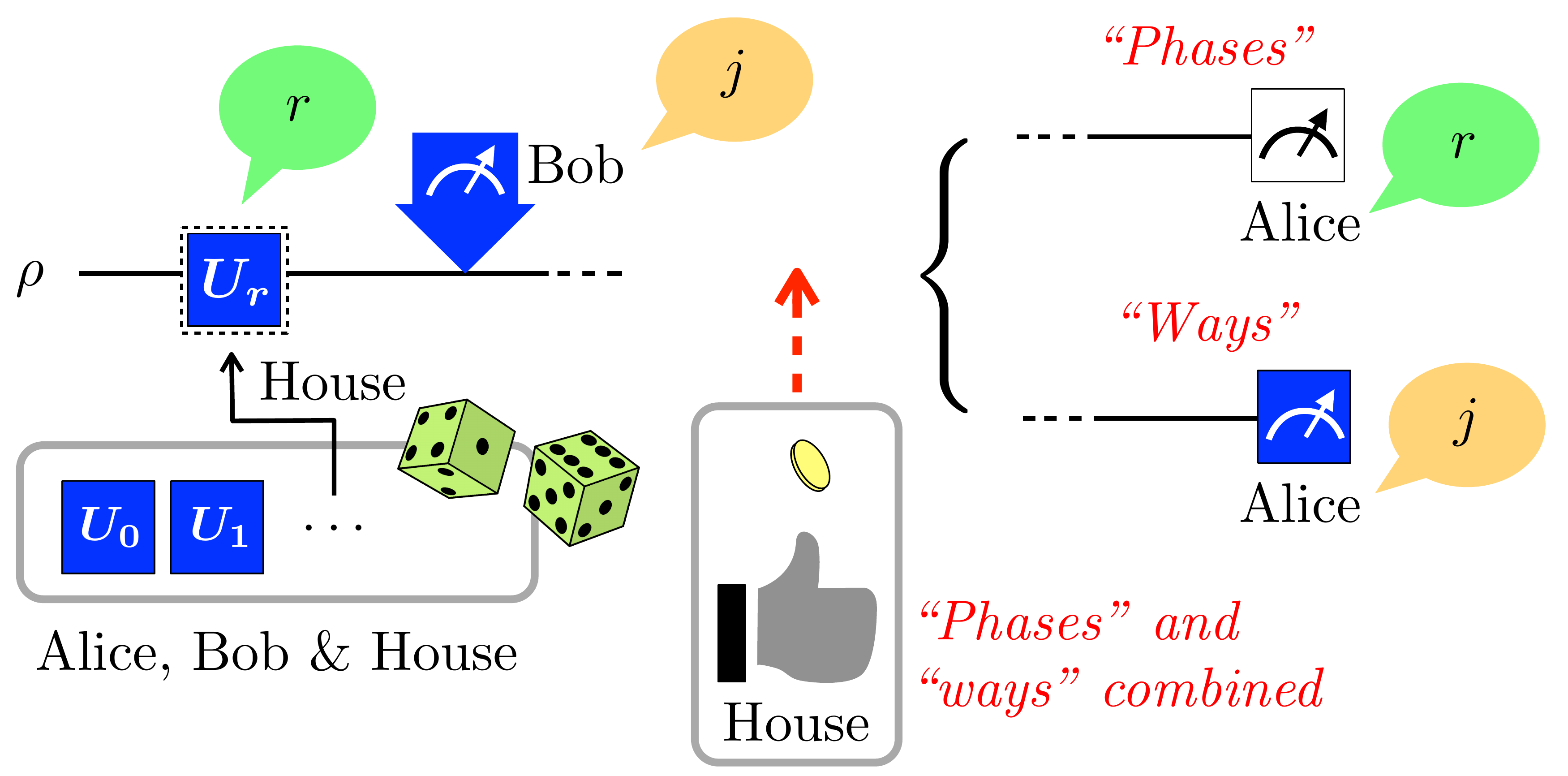}
\caption{Schematic representation of the games.  Operations that are basis dependent, as they meet requirements specific to the path basis, are colored blue.}
\label{f:1}
\end{center}
\end{figure}
%

%We consider the most general multi-port interferometric setup and put forward an hypothesis testing scenario which provides a duality relation between coherence and which-path information involving operationally defined quantities.

Before going into detail, let us introduce the games (see Fig.~\ref{f:1}) in their simplest terms. There are three parties, Alice and Bob, who cooperate to try to win, and the House, who supervises the games.  The games are played on an $n$-port interferometer fed with a single particle in a quantum state $\rho$, supplied by Alice and Bob.  The House can apply one of $n$ sets of phases to the paths in the interferometer, and each set is labelled.  These sets are known to Alice and Bob and each is equally likely.  The interferometer paths are coupled to path detectors.  The interaction between the particle and the detectors can be described by a controlled-unitary operation; if the particle is in a particular path, the detectors end up in a state that depends on that path~\cite{englert}.  Alice is given access to the particle and Bob is given access to the detectors, and once they are, they cannot communicate.  The House then flips a fair coin to decide which of the two games Alice and Bob will play.  In the game called \emph{phases}, Alice tries to determine the label corresponding to the set of phases that has been applied to the interferometer paths, and Bob does not play.  They win if Alice guesses the correct label.  In the game called \emph{ways}, Alice measures the particle in the path basis, and Bob measures the detectors, and they win if their measurement results agree.  We are interested in the winning probability of the combined game.

In more detail, we can describe a  general (single-particle) multi-path interferometer in terms of a special orthonormal basis
 $\{\ket{j}\}_{j=0}^{n-1}$ of ${\mathbb C}^n$, where~$\ket{j}$ is the state of the particle when it takes the
 $j$-th path. We will refer to it as the path basis.  We shall denote the sets of phases as $\phi_r:=\{\phi_r^j\}_{j=0}^{n-1}$, $r=0,1,\dots, n-1$, and the operator that applies them to the particle in the interferometer is
 \begin{equation}
U_r=\sum_{j=0}^{n-1}{\rm e}^{i\phi_r^j}|j\rangle\langle j| . 
\label{boxes}
\end{equation}
An $n$-path interferometer can perfectly discriminate among at most $n$ different such sets of phases.  The interaction of the particle in the interferometer with the detectors is described by a controlled unitary operation $|j\rangle|0\rangle \rightarrow |j\rangle |\eta_{j}\rangle$, where $|0\rangle$ is the initial state of the detectors, and $|\eta_{j}\rangle$ is the state of the detectors if the particle is in path $j$.  If the initial state of the particle was $\rho = \sum_{j,k=0}^{n-1}\rho_{jk} |j\rangle\langle k|$, the state after the phases and the controlled unitary have been applied is
\begin{equation}
\rho^{\rm AB}_r=\sum_{j,k=0}^{n-1}{\rm e}^{i(\phi_r^j-\phi_r^k)}\rho_{jk}\; |j\rangle\langle k|\otimes|\eta_j\rangle\langle \eta_k |,
\label{rhoAB}
\end{equation}
where the superscript $\rm AB$ indicates that this is the bipartite state shared by Alice and Bob.

%%%%%%%%%%%%%%%%%%%%%%%%%%%%%%%%

% We would like to mention that the above description includes the most general interferometric experiment where (Alice) aims at discriminating $n$ hypothesis encoded as relative phases between the paths andÊÊsimultaneouslyÊÊ(Bob) aims at discriminating which of the $n$ paths the particle takes \footnote{Note thatÊÊin an interferometer of $n$ paths the maximum number of hypothesis that can be perfectly encoded is precisely $n$.} Indeed, in our setting Alice is not limited to a particular way of recombining the paths, but may use the most general measurement to figure out the encoded phase. In addition,ÊÊBob is allowed to find out the path of the particle by making use of the most general quantum instrument, as long as it does not change the particle's path, or rather, the probability of the particle being in any particular path.ÊÊWe hence put forward that,Êin the context of the wave-particle duality, the \emph{which-path measurement} must map~$| j\rangle\langle j|$ into itself for any path $j$. This implies that Bob's measurement is a so called {\em genuinely incoherent operation}, or GIO~\cite{streltsov}, on $\rho$, which cannot create coherence. One can easily show that these operations can be accomplished by Bob appending an ancilla, which in this version replaces the detectors, in the state $|0\rangle$ to the particle and applying a controled-unitary map $|j\rangle |0\rangle \rightarrow |j\rangle |\eta_{j}\rangle$ to the combined system.ÊÊThe result is again $\rho^{\rm AB}_r$.

We would like to mention that one can take a more resource-theoretically-inspired approach to this problem.  In this scenario, Bob is given access to the particle itself. The important resource here is the coherence of~$\rho$ with respect to the path basis. %after the phases have been applied rather than access to the detectors.  
Bob then performs a measurement in order to determine which path the particle took.  
For this purpose he is allowed to make use of the most general quantum instrument~\cite{Davis}, as long as it does not change the particle's path, or rather, the probability of the particle being in any particular path. We thus put forward the notion of ``path measurement'' as a quantum instrument that maps~$| j\rangle\langle j|$ %(the state where the particle is in the $j$-th path) 
into itself for any path~$j$. This implies that Bob's measurement is a so called {\em genuinely incoherent operation}~\cite{streltsov}%, on~$\rho$
, which cannot increase the coherence of $\rho$. One can easily show that these operations can be accomplished by Bob appending an ancilla, which in this version replaces the detectors, in the state $|0\rangle$ to the particle and applying a controled-unitary map
$|j\rangle |0\rangle \rightarrow |j\rangle |\eta_{j}\rangle$ to the combined system.  The result is again $\rho^{\rm AB}_r$. %Note that Bob can do the above independently of the House's choice of phases, as  %what is the House's choice for the black box. 
%their application to the paths commutes with Bob's measurement.

%In order to obtain a duality relation via the game, however, it is necessary to restrict the kind measurement Bob performs, otherwise and he and Alice can win the game with probability one (see the supplemental material).  In particular, Bob's measurement must map $|j\rangle\langle j|$ into itself. This implies that Bob's measurement maps the set of incoherent states into itself, so that Bob's measurement cannot create coherence.  This can be accomplished by Bob appending an ancilla, which in this version replaces the detectors, in the state $|0\rangle$ to the particle and applying a map
%$|j\rangle |0\rangle \rightarrow |j\rangle |\eta_{j}\rangle$ to the combined system.  The result is again $\rho^{\rm AB}_r$.

Some comments on this version of the game are in order. First, note that Bob can do the above independently of the House's choice of phases, as  %what is the House's choice for the black box. 
their application to the paths commutes with Bob's measurement; %Note also that he can delay the last measurement  step until after Alice has performed her measurement. This has no effect on Alice's ability to play the game.  
the operations defined in Eq.~(\ref{boxes}) are incoherent with respect to the path basis, just as Bob's measurement is.
Second, if no restrictions on Bob's measurement were imposed, %he could sneak information about his outcome in the state of the particle (e.g., by changing the probabilities of Alice's outcomes%when they play {\em ways}
%). This renders the games trivial. In the supplemental material we show that Bob 
he could always  {\em perfectly} correlate his outcomes with Alice's, i.e., they could always win {\em ways} with probability one, without compromising Alice's ability  to win {\em phases}. %with a probability limited only by the coherence of the input state. 
In particular, if the input state were maximally coherent, they could win with probability one, which would render the game trivial. No restrictions on the winning probabilities, and thus no duality relation, would exist in this situation (see supplemental material for more details).% in such  sgive an explicit example where Alice and Bob win with probability one.  So, we impose the rule that  BobÕs measurement cannot change the probability of finding the particle in any of the $n$ paths of the interferometer. 

%%%%%%%%%%%%%%%%%%%%%%%%%%%%%%%%%%%%%%

If Alice and Bob play \emph{ways}, Alice will measure the particle in the path basis.  Bob then knows that the detectors are in one of the states in the ensemble 
$
\left\{ \rho_{jj},|\eta_j\rangle\langle \eta_j|\right\}_{l=0}^{n-1}
$.
Thus, the best he can do is to perform the optimal measurement that discriminates among the pure states $\{|\eta_j\rangle\}_{j=0}^{n-1}$, each with probability given by $\rho_{jj}$. If Alice and Bob play \emph{phases}, Bob does nothing (in fact, nothing Bob does affects Alice's ability to win phases), and Alice is faced with the task of determining the label of the set of phases that was applied by discriminating among the states
\begin{equation}
\rho_r=\sum_{j,k=0}^{n-1}{\rm e}^{i(\phi_r^j-\phi_r^k)}\rho_{jk}\langle\eta_k|\eta_j\rangle \; |j\rangle\langle k| ,
\label{rhos}
\end{equation}
each of which has a probability of $1/n$. There is no restriction on the measurement Alice can perform to accomplish her task.  We denote Alice's success probability by $P_{\rm ph}$.  It was shown in~\cite{pianiadessopra} that its maximum value, $P^*_{\rm ph}$, is achieved with the choice~$\phi_r^*=\{2\pi r j/n\}_{j=0}^{n-1}$, in which case $P^*_{\rm ph}=X+1/n$, where
\begin{equation}
X:={1\over n}{\cal C}_{\ell_1}(\rho_r)={1\over n}\sum_{\two{j,k=0}{j\not=k}}^{n-1}\left|\rho_{jk}\langle\eta_k|\eta_j\rangle\right|.
\label{def X}
\end{equation}
Here ${\cal C}_{\ell_1}(\rho_r)$ is the $\ell_1$ coherence measure of any of the states in Eq.~(\ref{rhos}).  So we have the bound
\begin{equation}
P_{\rm ph}\le X+{1\over n}.
\label{Pph < X}
\end{equation}
We will next derive our first duality bound, which we present in the form of a theorem 
\begin{theo}\label{t:1}
The maximum probability that Alice and Bob win the combined game is
\begin{equation}
P_{\rm win}:={1\over2}\left(P_{\rm ph}+P_{\rm way}\right)\le {1\over2}+{1\over2\sqrt n},
\label{P_win}
\end{equation}
where $P_{\rm way}$ is the probability that Alice and Bob win if they play {\em ways}.
This bound can be attained only if the input state $\rho$ is maximally coherent, namely, if
 $\rho=|\psi\rangle\langle \psi|$, where $|\psi\rangle=(1/\sqrt{n}\,)\sum_{j=0}^{n-1}|j\rangle$.  In this case, it is sufficient that the phases are given by $\phi_r^*$, %$\phi_r^j=2\pi sj/n$, 
 and~$\{|\eta_j\rangle\}_{j=0}^{n-1}$ are   symmetric states with constant overlap,
\begin{equation}
\langle\eta_j|\eta_k\rangle={1\over2}+{1\over2+2\sqrt n},\ 0\le j,k\le n-1,\ j\not=k. %,\quad p_j={1\over n},
\label{overlaps&p}
\end{equation}
%
%attain this maximum winning probability; i.e.,  they are optimal.
\end{theo}

\noindent To derive this result, we use the following lemma, which we prove in the supplemental material:

\begin{lemma} \label{l:1}
For any set of states~$\{\rho_r^{\rm AB}\}_{r=0}^{n-1}$ of the form of Eq.~(\ref{rhoAB}), the following bound holds:
\begin{equation}
X- {n-2\over n}(1-P_{\rm d})-{2\over n}\sqrt{(n-1)P_{\rm d}
(1-P_{\rm d})}  \le 0 ,
\label{bound}
\end{equation}
where $X$, Eq.~(\ref{def X}), is the normalized $\ell_1$ coherence measure of any of the states~${\rm tr}_{\rm B}\!\left(\rho_r^{\rm AB}\right)$ in Eq.~(\ref{rhos}), and~$P_{\rm d}$ is the maximum success probability of discrimination among the states, %after Alice's measurement in the computational bases $\{|j\rangle_{\rm A}\}_{j=0}^{n-1}$ has been performed. 
$\{|\eta_j\rangle\}_{j=0}^{n-1}$.
The bound is attained if \mbox{$\langle\eta_j|\eta_k\rangle=s$}, $0\le j,k\le n-1 $, $j\not=k$, with $s$ independent of $j$ and~$k$, and~$\rho_{jk}=1/n$.
\end{lemma}

%Assuming the coin is unbiased, 
\noindent The wining probability is $P_{\rm win}=(P_{\rm ph} + P_{\rm d})/2$, where $P_{\rm ph}$ ($P_{\rm way}=P_{\rm d}$) is the probability that Alice (Bob) guesses the phase (the state~$|\eta_j\rangle$) right. %From~\cite{pianiadessoprl} we have that the former is related to the $\ell_1$-norm of coherence through $P_\phi=X+1/n$. 
Therefore, as can we read off from Eqs.~(\ref{Pph < X}) and~(\ref{bound}), one has 
\begin{equation}
P_{\rm win}\le{1\over2}-{1\over2n}+{P_{\rm d}+\sqrt{(n-1)P_{\rm d}
(1-P_{\rm d})} \over n}.
\end{equation}
The right hand site of this inequality can be easily maximized over $0\le P_{\rm d}\le 1$, yielding Eq.~(\ref{P_win}) at the value~of
\begin{equation}
P_{\rm d}=P_{\rm ph}={1\over2}+{1\over2\sqrt n}.
\label{max Ps}
\end{equation}
For  an ensemble of equiprobable states, all of them with the same overlap $s$, the probability of success is known to be given by~\cite{englert2} (see the supplemental material for an alternative derivation)
\begin{equation}
P_{\rm d}=\left[{\sqrt{1+(n-1)s}+(n-1)\sqrt{1-s}\over n}\right]^2.
\end{equation}
One can easily check that the overlaps in Eq.~(\ref{overlaps&p}) yield Eq.~(\ref{max Ps}). This concludes the proof of Theorem~\ref{t:1}.

%\medskip

 Theorem~\ref{t:1} %Eq.~(\ref{P_win}) 
provides an operational meaning to  wave-particle duality. %Asymptotically, it becomes $P_{\rm d}\le 1/2$, so 
%Quantum mechanics sets a fundamental limitation by which discriminating the phases applied incoherently on the paths of %an interferometer  and guessing a partner's outcome of a measurement, where the two operations are performed in the %same chosen basis, cannot both be done perfectly with the same resource state. 
In the limit~$n\to\infty$, a choice of measurement for perfect path determination completely prevents Alice from discriminating phases. Likewise, from a setup for perfect phase discrimination no information can be inferred about the path followed by the particle in the interferometer. We can hence view our result as an ``uncertainty relation" for these two incompatible tasks. %There is a trade off between the amount of coherence required to accomplish the phase discrimination task and the amount of ``which-path" information required to guess a partner's outcome, i.e.,  to enable the correlation of Alice's and Bob's measurement outcomes. 
For finite $n$, the maximum winning probability is larger than~$1/2$, thus providing Alice and Bob with a means  to get  advantage over the House, if they make the right choices indicated in Theorem~\ref{t:1}. 

Note that the bound in Eq.~(\ref{P_win}) for $n>2$ cannot be derived from any of the quadratic wave-particle duality relation found in the literature. Take, e.g., the duality relation of reference~\cite{bagan} and recall Eq.~(\ref{Pph < X}). %again that~\mbox{$P_{\rm ph}\le X+1/n$}. 
Then, we obtain the much weaker bound~\mbox{$P_{\rm win}\le 1/\sqrt{2}+O(1/n)$}, where the $1/\sqrt2$ term results typically from quadratic bounds of the form of Eq.~(\ref{duality1}). This shows that this type of duality relation for multi-port interferometers is generically not tight.

We now turn to our strongest wave-particle duality relation. %, closer in spirit to our result in~\cite{bagan}. 
Let us first introduce the variables $x$ and~$y$, defined~as
\begin{equation}
x:={X\over1-1/n},\quad
y:={P_{\rm d}-1/n\over 1-1/n} .
\label{x & y}
\end{equation}
The variable $x$ is nothing but a normalized version of the $\ell_1$ coherence measure. The variable $y$ was introduced in~\cite{jaeger} and used in~\cite{coles1,coles2}. It takes into account that by random guessing, without performing any actual measurement,  the probability of identifying the path that a particle took in an $n$-port interferometer (i.e., $P_{\rm way}$ in the language of games) is~$P_{\rm d}=1/n$. %no information about the path a particle took in an $n$-port interferometer  is a measure of the path information 
These two variables take values in the interval $[0,1]$. 
With their aid, the duality bound of~\cite{bagan} can be written in a more suggestive form~as
\begin{equation}
x^2+y^2\le 1.
\label{circ}
\end{equation}
%
%where we have normalized both our $\ell_1$-norm of coherence and path information as 
This equation states that wave-particle duality constrains the value of $x$ and $y$ arising from any physically realizable interferometric experiment to lie inside the positive quadrant of a unit circle in the $x$-$y$ plane. Because of the inequality in Eq.~(\ref{Pph < X}), %the inequality $P_{\rm ph}\le X+1/n$, 
the very same region is obtained using the more symmetrical definitions
\begin{equation}
x:={P_{\rm ph}-1/n\over1-1/n},\quad
y:={P_{\rm way}-1/n\over 1-1/n} ,
\label{x & y op}
\end{equation}
which involves only operational quantities. Now we can state our strongest result in the form of a theorem
\begin{theo}\label{t:2}
All points $(x,y)$, where $x$ and $y$ are the variables defined in Eq.~(\ref{x & y}) or Eq.~(\ref{x & y op}), are constrained to lie on the region
%this region consists of the points $(x,y)$ that lie on 
${\mathscr T}\cup{\mathscr E}$, where $\mathscr T$ is the right triangle (orange region in Fig.~\ref{f:2}) $x+y-1\le 0$, $x>0$, $y>0$,  and~${\mathscr E}$ is the portion of ellipse (e.g., yellow region in~Fig.~\ref{f:2})
\begin{equation}
\left({x+y-{n-2\over n-1}\over {\sqrt{n}\over n-1}}\right)^2+\left({x-y\over \sqrt{n\over n-1}}\right)^2\le1,\quad x,y>0.
\label{ellipse}
\end{equation}
The conditions \mbox{$\langle\eta_j|\eta_k\rangle=s$}, $0\le j,k\le n-1 $, $j\not=k$,  with~$s$ independent of $j$ and~$k$, and~$\rho_{jk}=1/n$ (input state maximally coherent), define the elliptical section of the boundary.
\end{theo}

\noindent The theorem follows immediately from Eq.~(\ref{bound}), which can be expressed as %Expressed in the same variables, Eq.~(\ref{bound}) can be cast as
\begin{equation}
nx-(n-2)(1-y)\le 2\sqrt{(1-y)\left[1+(n-1)y\right]} .
\end{equation}
%
%one can see that this region consists of the points $(x,y)$ that lie on ${\mathscr T}\cup{\mathscr E}$, where $\mathscr T$ is the right triangle (orange region in Fig.~\ref{f:2}) $x+y-1\le 0$, $x>0$, $y>0$,  and ${\mathscr E}$ is the portion of ellipse (yellow region in~Fig.~\ref{f:2})
%%
%\begin{equation}
%\left({x+y-{n-2\over n-1}\over {\sqrt{n}\over n-1}}\right)^2+\left({x-y\over \sqrt{n\over n-1}}\right)^2\le1,\quad x,y>0.
%\end{equation}
%%

From Eq.~(\ref{ellipse}), one can read off the location of the center of the ellipse $\mathscr E$, its semi-major and semi-minor axis and its orientation. In particular, the center, $O$, and the semi-minor axis, $OB$, lie on the line $x=y$.  The semi-major axis $OA$ is parallel to the line $x+y=1$, at a distance of $[\sqrt2(n-1)]^{-1}$ below it. The length of the semi minor axis is $\overline{OB}=\sqrt{n/2}/(n-1)$. We readily see that as $n$ grows, $\overline{OB}$ becomes smaller and the semi-major axis approaches the line $x+y=1$. In the limit~\mbox{$n\to\infty$} the ellipse collapses to the line $x+y=1$ and the physical region becomes the orange triangle in Fig.~\ref{f:2}; the duality relation becomes linear, $x+y\le1$. Theorem~\ref{t:1} defines the section $x=y$ of the physical region.
%%
%\begin{equation}
%x+y=1.
%\end{equation}
%%

We would like to emphasize that the duality relation defined by these regions is tight. Theorem~\ref{t:2} identifies input states and overlaps of Bob's (the detector) states $\{|\eta_j\rangle\}_{j=0}^{n-1}$ that characterize the boundary. In particular, only if the interferometer is fed with a maximally coherent state, $|\psi\rangle=(1/\sqrt n)\sum_{j=0}^{n-1}|j\rangle$, can $x$ and $y$ lie on the boundary.

\emph{Acknowledgment}. This publication was made possible through the support of a Grant from the John Templeton Foundation. The opinions expressed in this publication are those of the authors and do not necessarily reflect the views of the John Templeton Foundation. Partial financial support by a Grant from PSC-CUNY is also gratefully acknowledged. The research of EB and JC was supported by the Spanish MICINN, through contracts FIS2013-40627-P and FIS2016-80681-P, the Generalitat de
Catalunya CIRIT, contract  2014SGR-966, and ERDF: European Regional Development Fund. EB also thanks Hunter College for the hospitality extended to him during his research stay.

%{\color{green} \medskip what follows should be in the supplemental material.}

\section{Proof of Lemma~\ref{t:1}}
%We assume that the state of the particle and detectors, when the particle is inside the interferometer, is
%
%{\color{red}Any bipartite pure state $|\Psi\rangle$ can be written as
%%
%\begin{equation}
%|\Psi\rangle = \sum_{j=1}^{N} \sqrt{p_{j}} |j\rangle |\eta_{j}\rangle 
%\end{equation}
%%
%with no loss of generality. Here $\{ |j\rangle\}_{j=0}^{N-1}$ is Alice's computational basis, on which the black box will apply the phases incoherently. The coefficients $p_j$ are properly normalized probabilities. 

The quantity, $\rho_{ii}$ is the probability that Alice will obtain the $i$-th outcome %when, if directed by the House, she performs a measurement in the computational basis.  
if she performs a measurement in the computational basis. Accordingly, $ |\eta_i\rangle$ is Bob's posterior state. The states $|\eta_j\rangle$ are normalized but non-orthogonal in general. If  Bob is asked to guess Alice's outcome, He will need to discriminate with maximum success probability what state of the ensemble~$\{\rho_{ii},|\eta_i\rangle\}$ Alice's measurement has produced. The optimal minimum-error measurement is described by a POVM (positive operator valued measure) defined by a set $\{\Pi_j\}_{j=0}^{n-1}$ of positive operators that sum up to the identity, $\sum_{j=0}^{n-1}\Pi_j=\openone$, and the success probability is given by
\begin{equation}
P_{\rm d}=\sum_{j=0}^{n-1} \rho_{jj} \langle\eta_j|\Pi_j|\eta_j\rangle .
\end{equation}
%
%are the path states, which are orthonormal, $p_{j}$ is the probability that the particle is in path $j$, and $\{ |\eta_{j}\rangle \, | j=1,2, \ldots N\}$ are detector states.  Each state $|\eta_{j}\rangle$ is a state of all $N$ path detectors, and these states are not assumed to be orthogonal but are assumed to be linearly independent.  In order to determine which path the particle is in, we need to discriminate the detector states $|\eta_{j}\rangle$.  These states are not orthogonal, and so they cannot be discriminated perfectly.  In order to discriminate them, we shall adopt the minimum-error strategy in which the probability of making a mistake is minimized.  
%
Let us first assume that $\{|\eta_j \rangle \}_{j=0}^{n-1}$ are linearly independent. Then, the POVM opretators are  one-dimensional projections, $\Pi_{j}=|m_{j}\rangle\langle m_{j}|$, for $j=0,1,2,\ldots n-1$, where the states $|m_{j}\rangle$ are orthonormal \cite{belavkin}.  The probability of successfully discriminating the states becomes
\begin{equation}
P_{\rm d}=\sum_{j=0}^{n-1} \rho_{jj} | \langle\eta_{j}|m_{j}\rangle |^{2} .
\end{equation}
It is convenient to express the success probability in a different form.  Let $|\tilde{\eta}_{j}\rangle = \sqrt{\rho_{jj}} |\eta_{j}\rangle$, and define the matrix~$B$ by
\begin{equation}
|\tilde{\eta}_{j}\rangle = \sum_{k=0}^{n-1} B_{kj}|m_{k}\rangle ,
\end{equation}
which implies that
\begin{equation}
P_{\rm d}=\sum_{j=0}^{n-1}|B_{jj}|^{2} .
\end{equation}
The matrix $B$ satisfies the relation $B^{\dagger}B = W$, where,
\begin{equation}
W_{ij}:=\langle\tilde\eta_i|\tilde\eta_j\rangle=\sqrt{\rho_{ii}\rho_{jj}}\langle \eta_i|\eta_j\rangle ,
\end{equation}
and $W$ is the (generalized) Gram matrix of the states~$|\tilde{\eta}_{j}\rangle$, $j=0,1,2,\ldots n-1$.
The maximum success probability of minimum error discrimination {\color{black} for any ensemble $\{\rho_{jj},|\eta_j\rangle\}$, i.e., for fixed $W$,}  is then given by
\begin{equation}
\label{Psucc}
P_{\rm d}=\hspace{-1em}\max_{\two{B~{\rm s.t.}}{B^\dagger\! B=W}}\hspace{-.0em}\sum_{i=0}^{n-1}|B_{ii}|^2 .
\end{equation}

For the coherence measure we use the  scaled version of the $\ell_1$-norm of coherence used in~\cite{baumgratz}, which can also be expressed in terms of $W$ as
\begin{equation}
X={1\over n}\sum_{\two{i,j}{i\not=j}}\left|\rho_{ij}\right|\left|\langle\eta_i|\eta_j\rangle\right|\le {1\over n}\sum_{\two{i,j}{i\not=j}}\left|W_{ij}\right| ,
\label{X def}
\end{equation}
where we have use that positivity ($\rho\ge0$) implies $|\rho_{ij}|\le\sqrt{\rho_{ii}}\sqrt{\rho_{jj}}$. Here and hereafter, the summation limits are omitted to keep the notation uncluttered. The bound is attained if $\rho_{ij}=1/n$ for all~$0\le i,j\le n-1$.
We note that the optimization in Eq.~(\ref{Psucc}) does {\em not} invalidate Eq.~(\ref{X def}), as the maximization is constrained by the condition $B^\dagger B=W$. So, hereafter we can assume that~$B$ maximizes the success probability for a given  {\color{black} Bob's ensemble $\{\rho_{ii},|\eta_i\rangle\}$. Keeping this in mind, the error probability is}
%%
%\begin{equation}
%\sum_{i,j}|B_{ij}|^2=1.
%\end{equation}
%%
%Thus
%%
\begin{equation}
1-P_{\rm d}= \sum_{\two{i,j}{i \neq j}} |B_{ij}|^2.
\label{1-Ps}
\end{equation}

We will show that the maximum value of $X$ depends in a very specific manner on the maximum probability of success $P_{\rm d}$.
We will find an upper bound to $X$, starting with Eq.~(\ref{X def}), which we can write as
\begin{equation}
X\le{1\over n}\sum_{\two{i,j}{i\not=j}}\left|\left(B^\dagger B\right)_{ij}\right|
={1\over n}\sum_{\two{i,j}{i\not=j}}\left|\sum_k B^*_{ki}B_{kj}\right|,
\label{starting point}
\end{equation}
and show that the bound is attained by (symmetric) states such that $\langle \eta_i|\eta_j\rangle=s$, $i\not=j$, and $\rho_{ij}=1/n$, where~$0\le s \le1$ is independent of $i$ and $j$. So, this class of states maximizes coherence for a given value of our  path information measure.

First, Eq.~(\ref{starting point}) can be bounded as
\begin{eqnarray}
X&\le&{1\over n}\hspace{-.25em}\sum_{\two{i,j}{i\not=j}}\hspace{-.5em}\sum_{\two{k}{k\not=i,j}}|B_{ki}||B_{kj}|\nonumber\\
&+&{2\over n}
\sum_i |B_{ii}|\sum_{\two{j}{j\not=i}}|B_{ij}|,
\end{eqnarray}
where we have used the triangle inequality. We next use Schwarz inequality in both terms to get
\begin{eqnarray}
X&\le& {1\over n}\hspace{-.25em}\sum_{\two{i,j}{i\not=j}}\hspace{-.25em}\sqrt{\sum_{\two{k}{k\not=i,j}}|B_{ki}|^2}
\sqrt{\sum_{\two{k}{k\not=i,j}}|B_{kj}|^2}\nonumber\\
&+&{2\over n}
\sqrt{\sum_i |B_{ii}|^2}\sqrt{\sum_i\Bigg(\hspace{-.5em}\sum_{\two{j}{j\not=i}}|B_{ij}|\Bigg)^2}.
\end{eqnarray}
The first terms can be written is a more transparent way introducing the notation $x=(i,j)$, $i\not=j$, and the vectors~{\boldmath${\cal V}$, ${\cal U}$} with components 
\begin{equation}
{\cal V}_x:=\sqrt{\sum_{\two{k}{k\not=i,j}}|B_{ki}|^2};\quad {\cal U}_x:=\sqrt{\sum_{\two{k}{k\not=i,j}}|B_{kj}|^2} ,
\end{equation}
as well as $\mbox{{\boldmath$v$}}^{(i)}$, $\mbox{{\boldmath$u$}}^{(i)}$, with components
\begin{equation}
v^{(i)}_j:=1, \quad u^{(i)}_j=|B_{ij}|,
\end{equation}
{\color{black} if $j\not=i$, and vanishing components otherwise.}
By recalling Eq.~(\ref{Psucc}), we have
\begin{equation}
X\le {1\over n}\sum_{x}{\cal V}_x \,{\cal U}_x
+{2\over n}
\sqrt{P_{\rm d}}\sqrt{\sum_i\Bigg(\sum_{\two{j}{j\not=i}}v^{(i)}_j u^{(i)}_j\Bigg)^2} .
\end{equation}
This suggests using again Schwarz inequality as
\begin{equation}
X\le {1\over n}\sum_{\two{i,j}{i\not=j}}\kern-.5em\sum_{\two{k}{k\not=i,j}}\kern-0.5em|B_{ki}|^2+{2\over n}\sqrt{(n-1)P_{\rm d}
\sum_{\two{i,j}{j\not=i}}|B_{ij}|^2
},
\label{bound -1}
\end{equation}
where we have used that
\begin{eqnarray}
&\displaystyle \|\mbox{\boldmath{${\cal V}$}}\|=\|\mbox{\boldmath{${\cal U}$}}\|=\sqrt{\sum_{\two{i,j}{i\not=j}}\kern-.5em\sum_{\two{k}{k\not=i,j}}|B_{ki}|^2};\\
&\displaystyle  \|\mbox{\boldmath{$v$}}^{(i)}\|=\sqrt{n-1};\quad
 \|\mbox{\boldmath{$u$}}^{(i)}\|=\sqrt{\sum_{\two{j}{j\not=i}}|B_{ij}|^2}. &
\end{eqnarray}
We note that the first term in (\ref{bound -1}) contains {\em all} the entries~$|B_{ij}|$, $i\not=j$, several times. Furthermore, by symmetry, each of these entries must appear the {\em same} number of times. Therefore, there must be an integer constant~$A_n$ such that
\begin{equation}
\sum_{\two{i,j}{i\not=j}}\kern-.5em\sum_{\two{k}{k\not=i,j}}|B_{ki}|^2=A_n \sum_{\two{i,j}{i\not=j}}|B_{ij}|^2 .
\end{equation}
We can compute $A_n$ by setting $|B_{ij}|=1$. This gives us $n(n-1)(n-2)=n(n-1)A_n$. Thus $A_n=n-2$, and
\begin{equation}
X\le {n-2\over n}\sum_{\two{i,j}{i\not=j}} |B_{ij}|^2+{2\over n}\sqrt{(n-1)P_{\rm d}
\sum_{\two{i,j}{j\not=i}}|B_{ij}|^2} .
\end{equation}
Recalling~(\ref{1-Ps}), we finally get
\begin{equation}
X\le {n-2\over n}(1-P_{\rm d})+{2\over n}\sqrt{(n-1)P_{\rm d}
(1-P_{\rm d})}  .
\end{equation}

One can immediately check that this bound is attained by the choice 
\begin{equation}
B_{ii}=\bb;\qquad B_{ij}=\w,\quad i\not=j   .
\end{equation}
The maximum probability of success is given by 
\begin{equation}
P_{\rm d}=n\bb^2 .
\label{Ps b}
\end{equation}
(since {\em all} the diagonal terms are equal, we know that the equation above gives indeed the optimal probability of success (see, e.g., Theorem 1 in~\cite{Ch Po}). Also,
\begin{equation}
W_{ii}=%\sum_{k}B^*_{ki}B_{ki}=
|B_{ii}|^2+\sum_{\two{k}{k\not=i}}|B_{ki}|^2=\bb^2+(n-1)\w^2={1\over n}.
\label{Wii}
\end{equation}
%%So, 
%$$
%p_i=W_{ii}=\bb'=1/n .
%$$
For $i\not=j$,
\begin{equation}
{\langle\eta_i|\eta_j\rangle\over n}=W_{ij}=2\w\bb+(n-2)\w^2:={s\over n}
\label{Wij}
\end{equation}
Thus, all the overlaps are equal. 

This proves Theorem~\ref{t:1} for independent $\{|\eta_j\rangle\}$. For linearly dependent states, the Gram matrix formalism cannot be applied since the POVM elements have in general rank greater than one. However, $P_{\rm d}$ and $X$ are both continuous functions of the components of the states $\{|\eta_j\rangle\}_{j=0}^{n-1}$ for any value of~$n$. This is apparent for $X$, since it just depends on all the overlaps of these states, as shown in Eq.~(\ref{X def}). As for the continuity of $P_{\rm d}$, %we first note that for any set~$\{|\eta_j\rangle\}_{j=0}^{n-1}$ and any finite value of $n$, the maximum success probability is a well defined quantity. Next, 
we can use the following lemma 

\begin{lemma} 
\label{l:2}
The maximum success probability of discrimination among the various elements of an ensemble of states~$\rho:=\{p_i,\rho_i\}_{i=0}^{n-1}$ is a  Lipschitz continuous function of the entries of the density matrices in the ensemble.

\end{lemma}

Let us prove Lemma~\ref{l:2}. Denote the maximum success probability of the lemma by $P_{\rm d}^{\rho}$. Let~$\| \rho \|$ be the ``ensemble norm", defined by
\begin{equation}
\|\rho\|:=\sqrt{\sum_{i=0}^{n-1} \|\rho_i\|^2_2}=\sqrt{\sum_{i,j,k=0}^{n-1}\left|\left(\rho_i\right)_{jk}\right|^2}.
\end{equation}
(Here, we consider the probabilities $p_i$ to be fixed quantities.) Then, for any two ensembles $\rho$ and $\sigma$, 
\begin{equation}
\left|P_{\rm d}^\rho\!-\! P_{\rm d}^\sigma\right|=
\left| \sum_{i=0}^{n-1}p_i{\rm tr}\!\left( \Pi^{\rho}_i\rho_i\right)\!-\!\sum_{i=0}^{n-1}p_i{\rm tr}\!\left( \Pi^{\sigma}_i\sigma_i\right)\right|.
\label{trs}
\end{equation}
%
%for any ensemble $\sigma=\left\{p_i,\sigma_i\right\}_{i=0}^{n-1}$ such that $\|\rho-\sigma\|<\delta$. To show that this statement is true, 
Note that the right hand side of Eq.~(\ref{trs}) is upper bounded by
\begin{eqnarray}
%\left|P_{\rm d}^\rho\!-\! P_{\rm d}^\sigma\right|\!\le\!
 \max\!\left\{
\sum_{i=0}^{n-1}p_i{\rm tr}\!\left[ \Pi^{\rho}_i\!\left(\rho_i\!-\!\sigma_i\right)\right],-\!\!
\sum_{i=0}^{n-1}p_i{\rm tr}\!\left[ \Pi^{\sigma}_i\!\left(\rho_i\!-\!\sigma_i\right)\right]\!
\right\}\!.
\end{eqnarray}
The two traces in brackets are, in turn, upper bounded by $k \|\rho-\sigma\|$, where $k$ is some constant (independent of~$\rho$ and~$\sigma$), since the operators $\left\{\Pi_i^\rho\right\}$ and $\left\{\Pi_i^\sigma\right\}$ are all bounded.
Thus, 
\begin{equation}
\left|P_{\rm d}^\rho\!-\! P_{\rm d}^\sigma\right|\le k \|\rho-\sigma\| ,
\end{equation}
which concludes the proof.
%%
%\begin{equation}
%k=\max\left\{  p_i |(\Pi_i^\rho)_{jk}| , p_i |(\Pi_i^\sigma)_{jk}|   \right\}
%\end{equation}
%%

Assume now that there exists some linearly dependent ensemble of states $\{\rho_{ii},|\bar \eta_i\rangle\}$ that violates Eq.~(\ref{bound}). So, the left hand side of this equation would yield a value~$\gamma_0>0$  for such ensemble. We can now change some components of the states $|\bar\eta_i\rangle$ by any (sufficiently  small) amount~$\epsilon>0$ and make them linearly independent. For the modified states, the left hand side of Eq.~(\ref{bound}) would now yield~$\gamma(\epsilon)\le 0$. But the expression on the  left hand side of Eq.~(\ref{bound}) is a continuous function of~$X$ and~$P_{\rm d}$, and thus, a continuous function of the components of the states that belong to the ensemble for any~$\epsilon$, and taking the limit~$\epsilon\to0$ we recover the original ensemble $\{\rho_{ii},|\bar \eta_i\rangle\}$, hence $\gamma_0=\lim_{\epsilon\to0}\gamma(\epsilon)\le0$ in contradiction with the assumption $\gamma_0>0$. In summary, Eq.~(\ref{bound}) holds for any ensemble $\{\rho_{ii},| \eta_i\rangle\}$, regardless whether the states are linearly dependent or independent.

\medskip

Before clossing this section, we note that one can easily solve for $b$ and $w$ the rightmost equations in Eqs.~(\ref{Wii}) and~(\ref{Wij}). Substituting the solution for $b$ in~Eq.~(\ref{Ps b}) one obtains
\begin{equation}
P_{\rm d}=\left[{\sqrt{1+(n-1)s}+(n-1)\sqrt{1-s}\over n}\right]^2.
\end{equation}

\section{\boldmath No restrictions on Bob's measurement  implies no duality relation: $P_{\rm way}=1$ for all input states}

We will show that if Bob is allowed to perform any measurement, not necessarily non-demolishing in the path basis,  then Alice and Bob can always win the \emph{ways} game, without affecting the success probability of the \emph{phases} game.

%If Bob is playing \emph{ways}, his 
Bob's task is to attain maximal correlations
between his outcomes and those that Alice will obtain when they play \emph{ways} and she is asked to measure the system in the path basis.
Bob's ability to determine which path the particle took is quantified by the correlation 
%The which-path information gained by Bob is quantified by the correlation measure
%
\begin{equation}
P_{\rm way}=\sum_{k=0}^{n-1} p(k,k),
\end{equation}
where $p(j,k)$ is the probability of Bob's outcome being~$j$ and Alice's $k$. In other words, the probability of winning this game, $P_{\rm way}$, is the probability of Bob guessing correctly the outcome of Alice's projective measurement on the path basis. Here, it is important to recall that Bob's measurement is a generalized measurement performed on the very same system that is subsequently measured by Alice.

With this in mind, Alice and Bob can devise the following strategy. Suppose that the sets of phases are given by $\phi_r^*= \{ (2\pi/n) r j\}_{j=0}^{n-1}$, with $r=0, 1, \ldots n-1$. Bob performs optimal minimum error discrimination of the phases, which he can do with success probability $P_{\rm ph}=(1+{\cal C}_{\cal R})/n$, where ${\cal C}_{\cal R}$ is the robustness of coherence of the input state $\rho$.
%For any input state $\rho$ and any set of black boxes $\{U_{r}\}_{r=0}^{n-1}$,  Bob can devise a measurement that allows him to guess with maximum success probability, $P^{*}_{\phi}\leq X+\frac{1}{n}$, the unitary implemented by the House.
Then, upon measurement outcome~$r$, Bob prepares the system on state $\ket{r}$, i.e., he encodes the output of the phase-measurement in the path-basis. Alice's part is now extremely  simple: she just has to measure in the path basis, regardless of whether the House asks her to play {\em ways} or {\em phases}. Clearly, by doing so she will maximize the correlations with Bob, i.e., $P_{\rm way}=1$ when they play \emph{ways}. When they play \emph{phases}, she will provide the same (optimal) guess as Bob. Thus,
\begin{equation}
P_{\rm win}={1\over2}\left(1+{1+ {\cal C}_{\cal R}\over n}\right).
\end{equation}
In particular, if the input state $\rho$ is maximally coherent, $ {\cal C}_{\cal R}= {\cal C}_{\ell_1}=n-1$, thus
$P_{\rm win}=1$, which renders the game trivial. In summary, if Bob's (which-way) measurement can change $p(k)=\sum_j p(j,k)$, no wave-particle duality relation can be observed.


\begin{thebibliography}{99}
\bibitem{baumgratz} T.\ Baumgratz, M.\ Cramer, and M.\ B.\ Plenio, \prl {\bf 113}, 140401 (2014).
\bibitem{winter} A.\ Winter and D.\ Yang, Phys.\ Rev.\ Lett.\ {\bf 116}, 120404 (2016) .
\bibitem{yadin} B.~Yadin, J.~Ma, D.~Girolami, M.~Gu, and V.~Vedral, arXiv: 1512.02085 (2015).
\bibitem{chitambar} E.~Chitambar and G.~Gour, Phys.\ Rev.\ Lett.\ {\bf 117}, 030401 (2016).
\bibitem{marvian} I.~Marvian and R.~Spekkens, Phys.\ Rev.\ A {\bf 94}, 052324 (2016).
\bibitem{pianiadessoprl} C. Napoli, T. R. Bromley, M. Cianciaruso, M. Piani, N. Johnston, and G. Adesso, \prl {\bf 116}, 150502 (2016).
\bibitem{pianiadessopra} M. Piani, M. Cianciaruso, T. R. Bromley, C. Napoli, N. Johnston, and G. Adesso. \pra {\bf 93}, 042107, (2016).
\bibitem{streltsov} A.~Streltsov, G.~Adesso, and M.~Plenio, arXiv:1609.02439 (2016).
\bibitem{wootters} W.\ K.\ Wootters and W.\ H.\ Zurek, \prd {\bf 19}, 473 (1979).
\bibitem{greenberger} D.\ M.\ Greenberger and A.\ YaSin, Phys.\ Lett.\ A {\bf 128}, 391 (1988).
\bibitem{jaeger} G.\ Jaeger, A.\ Shimony, and L.\ Vaidman, \pra {\bf 51}, 54 (1995).
\bibitem{englert} B.-G.\ Englert, \prl {\bf 77}, 2154 (1996).
\bibitem{durr} S.\ D\"{u}rr, \pra {\bf 64}, 042113 (2001).
\bibitem{bimonte} G.\ Bimonte and R.\ Musto, J.\ Phys.\ A {\bf 36}, 11481 (2003).
\bibitem{englert2} B.-G.\ Englert and J. A. Bergou, Optics Communications {\bf 179}, 337 (2000).
\bibitem{jakob} M.\ Jakob and J.\ A.\ Bergou, \pra {\bf 76}, 052107 (2007).
\bibitem{englert3} B.-G.\ Englert, D.\ Kaszlikowski, L.\ C.\ Kwek, and W.\ H.\ Chee, International Journal of Quantum Information {\bf 6}, 129 (2008).
\bibitem{angelo} R.~M.~Angelo and A.~D.~Ribeiro, Found.\ Phys.\ {\bf 45}, 1407, (2015).
\bibitem{pati} M.\ N.\ Bera, T.\ Qureshi, M.\ A.\ Siddiqui, and A.\ K.\ Pati, \pra {\bf 92}, 012118 (2015).
\bibitem{bagan} E.~Bagan, J.~Bergou, S.~Cottrell, and M.~Hillery, Phys.\ Rev.\ Lett.\ {\bf 116}, 160406 (2016).
\bibitem{coles1} P.~Coles, J.~Kaniewski, and S.~Wehner, Nature Communications {\bf 5}, 5814 (2014)..
\bibitem{coles2} P.~Coles, Phys.\ Rev.\ A {\bf 93}, 062111 (2016).
\bibitem{Davis} E.~B.~Davies, J.~T.~Lewis, %ÒAn operational approach to quantum probability,Ó 
Comm. Math. Phys., {\bf 17}, 239~%-260, 1970.
(1970).
\bibitem{belavkin} V.~P.~Belavkin, Stochastics {\bf 1}, 315 (1975).
%\bibitem{englert4} B.-~G.~Englert and J.~\v{R}eha\v{c}ek, J.\ Mod.\ Optics {\bf 57}, 218 (2010). %{\color{red} I'm sure this expression was obtained before, but I could not find the reference. It's also in U. Herzog, \pra {\bf 86}, 032314 (2012) as a particular case when the inconclusive outcome happens with zero probability. }
\bibitem{nielsen} M.\ Nielsen and I.\ Chuang, \emph{Quantum Computation and Quantum Information} (Cambridge University Press, Cambridge, 2000).
%\bibitem{Herzog?} U. Herzog, \pra {\bf 86}, 032314 (2012). {\color{red} I'm sure this expression was obtained before, but I could not find the reference. It's certainly in this paper as a particular case when the inconclusive outcome happens with zero probability. }
\bibitem{Ch Po} G. Sentis, E. Bagan, J. Calsamiglia, G. Chiribella, and R. Munoz-Tapia, \prl {\bf 117}, 150502 (2016).
\end{thebibliography}
\end{document}